# Live Objects All The Way Down

## Removing the Barriers between Applications and Virtual Machines

Javier E. Pimás[a], Stefan Marr[b], and Diego Garbervetsky[a]

a   Departamento de Computación, Universidad de Buenos Aires, Argentina
b   School of Computing, University of Kent, United Kingdom

**Abstract**   Object-oriented languages often use virtual machines (VMs) that provide mechanisms such as just-in-time (JIT) compilation and garbage collection (GC). These VM components are typically implemented in a separate layer, isolating them from the application. While this approach brings the software engineering benefits of clear separation and decoupling, it introduces barriers for both understanding VM behavior and evolving the VM implementation. For example, the GC and JIT compiler are typically fixed at VM build time, limiting arbitrary adaptation at run time. Furthermore, because of this separation, the implementation of the VM cannot typically be inspected and debugged in the same way as application code, enshrining a distinction in easy-to-work-with application and hard-to-work-with VM code. These characteristics pose a barrier for application developers to understand the engine on top of which their own code runs, and fosters a knowledge gap that prevents application developers to change the VM.

We propose Live Metacircular Runtimes (LMRs) to overcome this problem. LMRs are language runtime systems that seamlessly integrate the VM into the application in live programming environments. Unlike classic metacircular approaches, we propose to completely remove the separation between application and VM. By systematically applying object-oriented design to VM components, we can build live runtime systems that are small and flexible enough, where VM engineers can benefit of live programming features such as short feedback loops, and application developers with fewer VM expertise can benefit of the stronger causal connections between their programs and the VM implementation.

To evaluate our proposal, we implemented Bee/LMR, a live VM for a Smalltalk-derivative environment in 22,057 lines of code. We analyze case studies on tuning the garbage collector, avoiding recompilations by the just-in-time compiler, and adding support to optimize code with vector instructions to demonstrate the trade-offs of extending exploratory programming to VM development in the context of an industrial application used in production. Based on the case studies, we illustrate how our approach facilitates the daily development work of a small team of application developers.

Our approach enables VM developers to gain access to live programming tools traditionally reserved for application developers, while application developers can interact with the VM and modify it using the high-level tools they use every day. Both application and VM developers can seamlessly inspect, debug, understand, and modify the different parts of the VM with shorter feedback loops and higher-level tools.

## The Art, Science, and Engineering of Programming

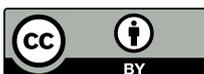

Removing the Barriers between Applications and Virtual Machines

## 1 Introduction

Over the last decades, language implementations enabled *Live Programming Environments (LPEs)*, allowing us to modify programs as they run. This feedback-driven exploratory programming style [25] lets application developers experiment with domain problems in short feedback loops with an immediate response to their actions.

Unfortunately, this exploratory programming style is restricted to application code, leaving the underlying language implementation unreachable and virtually immutable for application developers. This is because LPEs are implemented using virtual machines (VMs). This intentional architectural separation improves portability and limits interdependencies. The interfaces VMs provide are designed to hide implementation details and limit the ways an application can interact with a VM and customize it to its needs, in effect making it *invisible* to the application programmer.

This strict separation is however a double-edged sword. While it provides benefits, it intentionally keeps application developers unaware of design and implementation details, which prevents them from fully understanding performance pitfalls and restrictions of the system. This can lead to inefficient application code, perhaps causing too many garbage collections, or inadvertently triggering unnecessary recompilations. The strict separation also prevents developers from improving the VM and harnessing it in their application, e.g. by leveraging the just-in-time compiler or the graph tracing facilities of the garbage collector. As a result, the architecture of state-of-the-art VMs makes it hard for application developers to develop the most efficient code possible.

While numerous projects explored modern ways of implementing VMs [1, 5, 10, 16, 26, 35, 37], none designed the VMs in a way so that the code of the VM itself can be changed while it is executing, because VM components such as the just-in-time compiler and garbage collector are separated from the hosted language. Thus, the exploratory programming possible in LPEs is not available to VM engineers, who have to endure longer feedback loops.

Of course understanding a VM and its essential complexity is also a hurdle. However, for application developers much of the overall complexity is merely accidental and stems from the programming environment in which the VM is implemented. The differences between the languages, tools, and basic development processes available for VM and application development cause a significant learning curve.

In this paper, we demonstrate that it is possible to design a VM so that it can be changed and worked on with the same tools known to application developers by avoiding the architectural separation. We argue for a classic object-oriented design that enables live object-oriented programming, enabling us to construct a *Live Metacircular Runtime* (LMR). Based on the following case studies that are taken from the development of a large industrial application with 1.1 million lines of code, we show that application-level problems can be solved with application-level solution strategies, showing the benefit of less opaque and more flexible VMs. The case studies are:

**GCT** Garbage Collector Tuning for a particular use case.

**JITC** Changing a JIT compiler to avoid recompiling methods too frequently.

**CompO** Optimizing the application with vector instructions added to the compiler.

Our case studies were conducted with Bee Smalltalk [23, 24], which we use to implement the Bee LMR. Bee/LMR is a self-hosted VM runtime, written in Smalltalk. It replaced the classic Bee VM, which was written in C++ and assembly. Bee/LMR allows application developers to modify the virtual machine code at run time, as they would do with any other application-level code within the programming environment. It has been deployed to clients and is used daily for development of a simulation product in the oil and gas industry. While our focus for Bee/LMR was practicality and stability, Section 6.1 shows that the performance is not too far off a widely used Smalltalk VM and faster than Ruby and Python.

The key contributions of this work are:

- An evaluation of the trade-offs of removing the architectural distinction between VM and application using three case studies on a large industrial application. The key benefits are the gains of immediacy of LPEs and the explorability and malleability of LMRs.
- The design of a runtime system without the architectural separation of VM and application, which we call *Live Metacircular Runtimes*.
- The implementation of a self-hosted Smalltalk with a just-in-time compiler and garbage collector, where every part of the runtime can be changed at run time.

## 2 Background: Live Programming and the Architecture of Virtual Machines

This section begins with a short characterization of live programming environments and then gives a brief overview of the architectures provided by virtual machines. First, we discuss the classic layered architecture where the VM enforces a separation. Then, we look at variations and approaches that affect this separation. This section also introduces Bee Smalltalk, the system used for our case studies.

### 2.1 Exploratory Programming through Live Programming Environments

The focus of this work is shortening feedback loops in VM development. A *Live Programming Environment* is a set of software tools that allows developers to continuously change programs while they are running, allowing them to change program components without restarting them so that the development feedback loop is immediate. Such environments have a long history [25], with the work on SOAR being perhaps one of the earliest examples [32] of Live Programming Environments for object-oriented languages based on Smalltalk. Self was also designed with that same philosophy [8] and its Klein [33] implementation comes the closest to what we consider an LMR, as we discuss in Section 2.3 below. Other languages such as Python, JavaScript, and Ruby were not originally implemented as LPEs, in practice requiring program restarts to evaluate new code. However, with time have gained features and tools that shorten feedback loops and allow for exploratory programming. For example, IPython [22] gives developers a *computational notebooks*, which is similar to a read-eval-print with mostly immediate feedback. Since it became popular, it got renamed to Jupyter to indicate the support of a wide range of languages and is widely used for instance for data analysis tasks.

## 2.2 A Layered Architecture: Separating VM and Application Layers

As discussed in Section 1, virtual machines are designed to clearly separate the virtual machine from the application, providing a layered architecture. This gives application developers a concrete and fixed target, enabling them for instance to port an application to compatible VMs as long they only rely on interfaces and behaviors guaranteed in a specification such as the JVM [19] or ECMAScript [13].

Features such as garbage collection and just-in-time compilation are automatic and VM developers will go through great efforts to make them unobservable. However, a VM may also offer APIs that enables an application to interact with the VM and configure it for its needs. For instance, many VMs allow an application to trigger garbage collection. Other common APIs may give access for instance to run-time statistics such as memory use, time spent on garbage collection or just-in-time compilation. Often these APIs will be limited to avoid disclosing implementation details and thus, to minimize the risk that an application will depend on them.

Common VMs that use this layered architecture include the JVM, .NET's Common Language Runtime, and the ECMAScript-compatible JavaScript virtual machines, e.g., V8, SpiderMonkey, and JavaScriptCore. The common implementations of these VMs use a mixture of C/C++ and assembly code. This gives VM developers the control to reach the desired performance, in exchange for high development effort.

However, this two-layer design also has its problems, since features are provided intentionally as a black box to the application developer. This hides causal connections and prevents application developers from understanding and relying on implementation details. Thus, application developers cannot simply inspect the garbage collector to know under which conditions it triggers or the JIT compiler to know when a method is considered for compilation. While many of these VMs are open source, they are typically not written in the same language as the application, and are not part of the codebase, which prevents application developers to use their usual development tool to observe, analyze, reuse, and change these components. The immediate feedback they get from their tools with few exceptions ends at the VM layer.

## 2.3 Alternative Designs and Variations

Various projects have experimented with changes and variations of this two-layer architecture. Lisp's tower of interpreters [28] and Smalltalk's self-hosted Squeak [16] are classic examples, which to a certain degree achieve the goal of implementing a language in itself. More recent examples of this approach include PyPy [26], Rubinius [14], JikesRVM [2], Squawk [27], Maxine [35], and SubstrateVM [36]. The self-hosted VMs are mostly written in the same or a subset of the language they support. While Lisp's tower of interpreters has the ability to change the language at each level, the approach more widely used for VMs can be seen with OpenSmalltalkVM for Squeak and PyPy, which are translated to C and then compiled statically with Javier E. Pimás, Stefan Marr, and Diego Garbervetskya standard C compiler. A bootstrap VM compiles JikesRVM, Maxine, SubstrateVM directly into a native executable image instead of C code. This approach gives some leeway and enables Squeak for instance to provide simulation tools that allow it to execute the VM code the same as normal Smalltalk code, which makes it possible to use the same development tools for developing the VM as for the application. However this simulation is often around 1000x slower than the real VM [6]. Similarly, the PyPy implementation can be executed as normal Python code and developers can use their standard Python tools to work with it. The biggest drawback is the low performance of executing a metacircular VM on top of itself, which is not optimized and meant to be used in production by application programmers. Furthermore, these approaches still have the two-layer architecture discussed before, which clearly separates applications from the VMs.

We illustrate these designs in Figure 1. Figure 1a shows the two-layer architecture, while Figure 1b depicts self-hosted VMs that can run the VM code as an application.

While still maintaining a two-layer architecture, Fully Reflective Execution Environments (FREE) [9] aims to widen the APIs provided by VMs to enable the customization of ideally all aspects of a VM. This is achieved by designing metaobject protocols that can for instance change how method lookup is done, fields are accessed, or even how garbage collection and JIT compilation are done. As illustrated in Figure 1c, this however still maintains the architectural distinction and does not give developers access to the implementation of the VM itself, only providing more abilities for customization. Pinocchio [34] is another example for a Smalltalk VM where the interpreter itself can be adapted via a metaobject protocol.

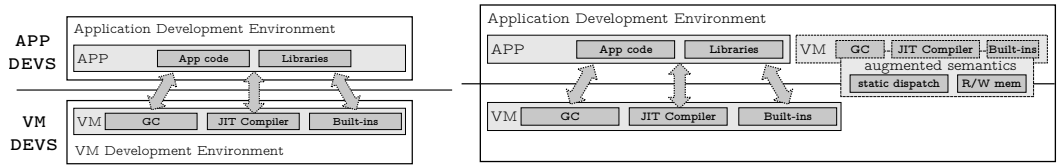

**(a)** Two-layer architecture of traditional VMs, clearly separating application and VM, with limited APIs for interaction.

**(b)** Self-hosted VMs maintain the layered architecture, but can allow developers to execute the VM as if it is an application.

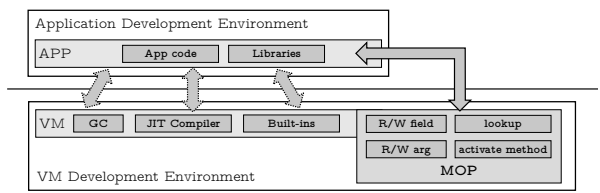

**(c)** VMs with Metaobject Protocols to customize field access, method lookup, and other aspects from within an application.

**Figure 1** VM implementation approaches with the traditional two-layer design.

**Beyond Object-Oriented Virtual Machines**    Outside the world of object-oriented VMs, projects like SML# [21] aim to combine application and runtime by compiling both into a static binary. Though, while the end product is a combination, SML# still separates the runtime from the application.

For general applications, Kiczales [17] proposed the notion of *Open Implementations*. In addition to a standard interface, a module is expected to provide a meta-interface. This meta-interface can then be used to control and adapt the implementation of a module, perhaps to select an algorithm that has better performance in a specific context. This idea builds on the notion of metaobject protocols [18] and metaarchitectures, in the context of which research explored a wide range of issues and designs [29], also including compile-time metaobject protocols [11]. Though, these approaches are what Fully Reflective Execution Environments [9] built on, which still relies on a fixed meta-interface, and thus, enforces a strong boundary between modules, or in our case VM and application.

**Designs with Reduced VM and Application Boundaries**    Klein [33] pushed the boundary a bit further. It was a self-hosted VM written in Self for Self, that was object-oriented, metacircular, and could be changed while running. In our understanding, it blurred the line between VM and application much more than any of the other approaches. Thus, it is an inspiration of our work, a first attempt to create a VM that does not strictly separate the application and can be evolved as one evolves an application. Unfortunately, Klein has neither implemented a self-hosted GC nor achieved the full Self functionality and performance.

Another point of inspiration is the Mist project,[1] which is explained as *a Smalltalk without a VM*. To our understanding, it also aimed at being a self-hosted Smalltalk implementation that compiles itself to native code, but is still at the early idea stage.

## 2.4  Bee Smalltalk

Bee Smalltalk, our research vehicle, initially ran on top of a traditional VM implemented in C++ and assembly. Bee/LMR was designed as a drop-in replacement for traditional VM and has a JIT compiler, garbage collector (GC), and numerous built-in functions. The main difference is that Bee/LMR is implemented in Smalltalk using a unified application/VM layer, while the traditional VM is split from the application and its C++ code is mostly invisible to application programmers.

Smalltalk methods are compiled into a bytecode format, and when first executed, are just-in-time compilation through a template JIT compiler. There is no interpreter mode. Bee/LMR has a generational GC, which is designed to keep objects valid at all time to facilitate a metacircular self-modifying runtime, as described in [23]. To ensure objects are valid, it uses object copying with external forward pointers, so that the GC only changes the mark bit in the object header. The overall design remained the same, but Bee/LMR now also adds a garbage-first GC [12] for the old heap.

---

[1] https://mist-project.org/, Martin McClure, 2016. Last access: 2023-10-27

New space consists of a big eden, and two smaller from and to spaces. When the eden is out of space, the generational scavenger is run. It moves most surviving objects in eden and from to to, and then flips from and to. Tenured objects are moved to the old zone, where they remain until the garbage-first collector is run. This is triggered by a heuristic that considers the heap growth since the last garbage-first collection. The old zone is divided in equally sized areas, and there is a large object zone for objects that exceed a size threshold. Tenured objects are bump-allocated in the old areas. As computation evolves some old objects become unreachable. The garbage-first collector keeps track of the reachable usage of each old area, where a low usage implies higher fragmentation. Before starting tracing, the GC selects areas to be evacuated, choosing the ones that have the lowest usage ratios. Evacuation makes objects allocated in fragmented areas become compacted into to other free areas.

The Smalltalk system itself comes with the common features of an image-based development environment. It uses as classic class browser with object inspectors and workspaces for arbitrary code evaluation. As for most Smalltalk systems, the debugger enables developers to inspect, explore, and modify the system at run time by changing state and code at will.

Bee is the platform for a simulation application in the oil and gas industry developed by a small team. Over time, the team realized that the strict separation between VM and application caused a too high cost, because the VM had to be maintained as a separate codebase. For example, investigating crashes no matter whether they were caused by the application or a VM bug took too much effort. As a small team with a production application of 1.1 million lines of code relying on Bee, it was decided that major steps had to be taken to make maintaining Bee easier for all team members. We describe the result in Section 4.

## 3  Application-Driven Case Studies

To motivate our exploration of a runtime design that removes the barriers between application and VM, we exemplify the trade-offs based on three case studies from the development of our application on top of Bee. Each case study describes a scenario and the solution steps needed in a state-of-the-art VM. We identify for each case study, the specific obstacles with these VMs and the features needed to enable application developers to benefit from their normal tools and short feedback loops.

### 3.1  Case Studies

For this study, we selected a case study on the performance impact of garbage collection, one on avoiding recompilations by the just-in-time compiler, as well as one on adding support for vector processor instructions for better performance.

#### 3.1.1  Case Study 1: Garbage Collection Tuning (GCT)
The allocation behavior of an application may trigger undesirable behavior by a garbage collector. For instance, too many allocations in a too small heap may cause too frequent collections, resulting in performance issues. Certain allocation patterns and heap structures may also cause long GC pauses. In some cases though more frequent collections can faster remove large numbers of temporary objects, which may make a smaller heap size have an overall better performance.

When diagnosing such issues in an application, we first need to identify when GC occurs, and collect statistics on heap size, object survival rates, and heap fragmentation. For long-running applications, we may also need to filter these statistics to specific parts, isolating the computation of interest.

A state-of-the-art VM may have a range of different collectors. For simplicity, we will only discuss a single-threaded scenario. Here a VM could have a generational GC that triggers when there is no space in the nursery to allocate new objects. VMs have various heuristics to trigger a minor or a full collection. An application may be able to use an API to read basic statistics or trigger a GC. However, changing the GC or fine-tuning its parameters, often requires to start the VM with different parameters. Full GC details may only be available via console logs enabled by command-line parameters or tooling interfaces that can be used via external tools to monitor GC and allocation details, as it is the case for JVMs. Thus, in most cases, an application developer has to drop out of the application development environment and possibly learn a different tool. All of today's solutions have in common that they maintain the strict separation between application and VM and rely on external mechanisms and tools to provide the desired insights.

For the specific performance problem at hand, after collecting the data, and analyzing it with new tools to be learned, or using custom code, we may find that one solution is to set a specific heap size while the relevant code is executing, which better balances the cost of GC and GC frequency. Though, adjusting the heap size mid-execution is not something generally supported by most VMs. However, modifying the VM is in the case of JVMs, JavaScript VMs, and many others is implausible for application developers. Not only will it require them to learn how to build these systems, but also to navigate possibly hundreds of thousands of lines of code.

In summary, for application developers, the available GC systems provide opaque interfaces that make it difficult to determine what happens and to gather GC statistics. Modifying a single line of code of the GC requires a significant initial cost for setup and building, and is a challenging task due to the switch to an external, opaque VM codebase, especially when written in a non-live language. Feedback loops are long, and details on how time is spent on memory-management routines are scarce. It is impractical to analyze GC triggering events, and the code of the GC can not be easily inspected or changed in the same way an application would be developed.

### 3.1.2 Case Study 2: Recurring Recompilations by the JIT Compiler (JITC)
Since run-time compilation comes at a cost, VMs use heuristics to decide when JIT compilation is worthwhile. However, as with all heuristics, unusual cases may result in undesirable performance.

In the case study, a product update got deployed to a customer, which reports that the new version is much slower than the previous one. The underlying problem is that in the new version, the application code dynamically adds and removes methods on an object, which invalidates JIT-compiled code causing frequent recompilations.

On a state-of-the-art VM, the developer would start to profile the application to detect the cause of the performance issues. However, since JIT compilation is transparent, a profiler normally does not show compilation time. It is also unlikely that an application developer would think to look for compilation statistics, or perhaps notice that the compiler thread may be indicated as busy for longer than usual.

Indeed, our application developer was not able to find the issue using the profiler. However, they constructed a reproducer that allowed them to use a binary search through the changes in the update to identify the cause, which led them to the code change that added the dynamic method updates. Thus, because of the strict separation between VM and application, the standard tools failed our developer, and they had to rely on the project using a suitable development approach to find the root cause.

After identifying the root cause, our developer needs a good understanding of the JIT compiler to devise a fix and avoid the recompilations. Changing the JIT compiler heuristics at fault is likely again too complex. Thus, they will need to rewrite the application code to avoid adding and removing methods dynamically.

### 3.1.3 Case Study 3: SIMD Optimizations (CompO)

Applications such as ours with a lot of floating-point arithmetics can gain performance by using vectorized operations in modern x86 and ARM processors. However, many high-level languages do not provide compiler optimizations or APIs to use them.[2]

To optimize the methods that do the computation, our application developers need to either directly or indirectly use the vectorized floating-point operations (SIMD: single instruction, multiple data). With state-of-the-art VMs, one could wait for or ask the VM vendor to add support for the vectorized operations. However, for instance for Java, it is likely still going to take another few years before support is finalized.

An alternative is to use an extension to the VM, that relies on low-level code, which makes the operations accessible. JVMs provide the Java Native Interface and other VMs have typically similar mechanisms. Another approach, exocompilation, [15] allows abstracting data processing algorithms from *scheduling*. This simplifies generating efficient machine code that targets varying hardware instructions and architectures.

However, these approaches require the application developer to switch the language and tools, which comes with extra effort and likely a longer-than-usual bug tail. In the worst case, there may not even be compiler intrinsics and our developer may need to use inline assembly. Depending on how the native interface works, it may also come with extra inefficiencies when switching between normal and extension code, perhaps because it needs to marshal high-level arguments into low-level arguments before computation, and possibly also to save VM state before calling native code.

---

[2] Even Java only has experimental support: https://openjdk.org/jeps/438. Last access: 2023-10-27

**Removing the Barriers between Applications and Virtual Machines**

## 3.2 Problems with State-of-the-Art VMs

The common theme in our case studies is that the strict separation between application and virtual machine for all its engineering benefits comes also at a cost.

### 3.2.1 Limited VM Observability (PG1)

Since VMs are designed to abstract and hide implementation details, directly available APIs are typically minimal. Even tooling interfaces as available for instance for JVMs may be limited by the desire to minimize run-time overhead when collecting data.

In our case studies, this meant that collecting the desired information about garbage collection either required to learn about external tooling or process log output. To understand the behavior of the just-in-time compiler, JVMs do provide the data as part of the tooling interfaces and the Java Management Extensions.[3] However, since it is transparent to the developer, not visible in profiles, and less commonly known as source for performance issue than GC, it is unlikely an application developer will consider it as a source of the issue.

### 3.2.2 Separate VM Development Mode (PG2)

Since the authors of this paper are VM developers, our first instinct is *yes, let us fix the VM*. However, in the common case, working on the VM is too different from working on the application to easily transfer language and tooling knowledge. In most cases, the VM is implemented in a more low-level language, has complex build steps that need to be understood, and requires the use of different development tools. As such, making a change to a VM requires a lot of additional learning before one can even start to explore the typical VM codebases of hundreds of thousands of lines of code. Thus, the standard solution for an application developer will be to find workarounds in the application code using the tools they know.

### 3.2.3 Long Edit-Compile-Run Feedback Loops for VM Components (PG3)

In cases like our case study were we want to use vectorized operations in our simulation code, dropping to the level of VM development is the only option. While building extensions is often supported by documentation and tooling that requires less learning than changing the VM itself, it is still a change of programming language and tooling and comes with the burden of longer edit-compile-run feedback cycles than what application developers are used to from their high-level languages.

With all this required learning, one can ideally change things, explore consequences, and have immediate feedback as application developers may be used to from their live programming environments. However, because extensions require lower-level languages, changes may require restarting the VM and application before taking effect, which is a drastic change in development flow. This lack of instant feedback increases the time from ideas to experiments and the cost of building such extensions.

---

[3] https://docs.oracle.com/en/java/javase/17/docs/api/java.management/java/lang/
management/CompilationMXBean.html Last access: 2023-10-27

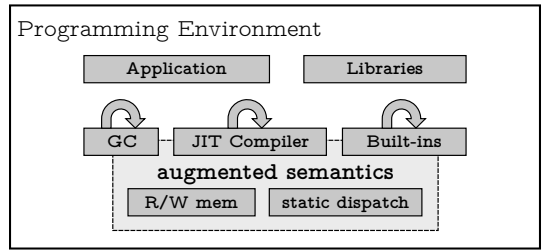

**Figure 2** Live Metacricular Runtimes provide VM components as standard object-oriented libraries in the same language as the application code.

### 3.3 Summary

Our three case studies show that the strict two-layer architecture separating VM and application code leads to knowledge and practical barriers. While the architecture ensures for instance portability, it limits the observability of VMs (PG1), typically leads to a separate development mode for the VM (PG2), and causes a longer edit-compile-run feedback loop for VM components than for application code (PG3).

## 4 Live Metacircular Runtimes

To overcome the limitations presented in the previous section, we propose Live Metacircular Runtimes (LMRs). LMRs enable a closer interaction between virtual machine and application. Instead of separating both into architectural layers, the VM becomes a set of object-oriented runtime components within the live programming environment. The components use standard object-oriented techniques to separate it from different subsystems in an application. However, by being integral part of the same system, LMRs synergize with and derive benefits from the advantages of live programming environments.

At its core, an LMR provides means to execute programs with an interpreter or JIT compilers, it provides built-in functions that implement low-level operations and it provides a garbage collector. An LMR seamlessly integrates those components as object-oriented libraries in the same language and same way as modularized application code. Therefore, all the tools used to program applications can be reused to program the VM itself. Figure 2 illustrates the high-level view of our LMR design.

LMRs allow programmers to harness the runtime in ways not possible in VMs that enfore a strict two-layer architecture. This is beneficial for VM developers when debugging and optimizing their code because it reduces feedback loops. Be removing the barrier between VM code and application code, they can use the same tools as application developers. Furthermore, application developers benefit from seeing the details of the VM when needed. If needed or desired, application developers can also more easily become VM developers and change the VM components with the same immediate feedback and using the tools they already know.

## 4.1 Bee Smalltalk's LMR Module

Bee Smalltalk incorporates a live metacircular runtime as a Smalltalk module called LMR, which implements JIT compilers for Smalltalk code, built-in functions written in Smalltalk, and both generational and full heap GCs. This LMR module replaced all functionality provided by the old C++ virtual machine, which is only needed for bootstrapping.

**Bootstrapping**   Since the Bee LMR module is Smalltalk code, it can be developed in standard Bee Smalltalk, either using the traditional VM or running on top of itself. It can also be loaded in other Smalltalks that can read the Bee Kernel and LMR modules. Since the LMR is made of just Smalltalk code, it can also be executed as a Smalltalk application, i.e., very similar to Squeak's simulation approach (see Section 2.3).

The bootstrap process generates a self-sustained image kernel with the core Smalltalk objects, the built-in functions, and the garbage collector. It also generates a library for the just-in-time compiler as execution engine. The result is written into a series of image segment files. A simple loader of around 1000 C++ lines of code is used to start the kernel which then loads the remaining components.

## 4.2 Augmenting the Bee Smalltalk Language to Simplify the LMR Implementation

The code of the LMR module is written in plain Smalltalk. As the language only supports late-bound dynamic message sends, the LMR module augments the traditional Smalltalk semantics with support for statically-bound message sends to enable low-level operations. This is done without changes in the syntax. As the LMR module contains the JIT compiler, which implements the semantics of the language, we can adapt them as needed. In Bee/LMR, we evolved our compiler to allow us to use metamessages to alter semantics of message sends. A metamessage is a message sent to an intermediate-representation (IR) node during compilation. The IR nodes implement metamessages that allow us to access object headers and other implementation-level concepts in an object-oriented way.

As mentioned, the main addition to Bee JIT compiler is support for statically bound message sends. These messages allow us to implement low-level operations such as reading and writing object headers, doing basic arithmetic, accessing memory unchecked, and changing control flow.

Listing 1 shows an example of the augmented semantics, connecting all components from the place of usage all the way to the implementation, reaching assembly generation. The method size sends a few metamessages: _isSmall, _smallSize and _largeSize. Metamessages are distinguished by a leading underscore and executed at compilation time. The implementation of a metamessage returns an IR node that represents the low-level operation being done.

■ **Listing 1** Use and implementation of augmented semantics.

```
1  ProtoObject >> size
2    ^self _isSmall ifTrue: [self _smallSize] ifFalse: [self _largeSize]
```

```
1  Node >> _isSmall
2    ^(self flags bitAnd: IsSmall asConstant) != 0 asConstant
```

```
1  Node >> flags
2    ^self loadByteAt: Flags
```

```
1  Node >> bitAnd: aNode
2    ^BitAndNode left: self right: aNode
```

```
1  Node >> _smallSize
2    ^self loadByteAt: SmallSize
```

### 4.3 Memory Management

The augmented semantics shown before are enough to allow implementation of a complete memory manager. The memory manager code is just plain Smalltalk code, and statically-bound methods are used to access object headers and convert between memory addresses and object references. Listing 2 shows how the `copyOf:` method integrates the access to a bit in the object header using the statically bound `_hasBeenSeen` method. Its implementation is given for the Node class. The `forward: object to: copy` method converts between addresses and object references, writes memory directly and marks an object as seen using other statically-bound methods.

Objects in Bee are created using bump pointer allocation. Bee implements a generational garbage collector [30, 31]. The young generation, or nursery, is collected using a copying scavenger. For the full heap collection, Bee implements a region-based one-pass opportunistic-only evacuating GC inspired by the garbage-first garbage collector [12, 38] and Immix [4].

### 4.4 Built-in Functions

In traditional Smalltalk systems, built-in functions consist of message lookup and *primitives*, and are shipped compiled as part of the VM. In Bee/LMR they are implemented in terms of augmented semantics and encapsulation. Listing 3 shows the implementation of SmallInteger»+ in the traditional VM, followed by the implementation of Bee/LMR. The <primitive> tag indicates this is a method is built into the VM.

**Removing the Barriers between Applications and Virtual Machines**

▪ **Listing 2** Example use and implementation of augmented semantics for the GC.

```
1  EdenCollector >> copyOf: anObject
2     ^anObject _hasBeenSeen
3        ifTrue: [self proxeeOf: anObject] ifFalse: [self doCopy: anObject]
```

```
1  EdenCollector >> forward: object to: copy
2     | index |
3     index := self forwardingIndexOf: object.
4     forwarders _asObject _basicAt: index put: copy.
5     object _beSeen
```

```
1  Node >> _hasBeenSeen
2     ^(self flags bitAnd: HasBeenSeen asConstant) != 0 asConstant
```

▪ **Listing 3** A primitive method in a traditional Smalltalk and in an LMR

```
1  SmallInteger >> + aNumber
2     <primitive: SmallIntegerPlus>
3     ^aNumber + self
```

```
1  SmallInteger >> + aNumber
2     | result |
3     aNumber _isSmallInteger ifFalse: [^aNumber + self].
4     result := self _smiPlus: aNumber.
5     ^result _overflowed
6        ifTrue: [self asLargeInteger + aNumber]
7        ifFalse: [result]
```

The addition primitive can fail if the argument is not an instance of SmallInteger or if the addition overflows. When such a failure happens, the primitive is aborted and the Smalltalk code that follows the tag is executed. For Bee/LMR, the same functionality of the primitive is written using metamessages, as explained in Section 4.2.

Finally, the method lookup mechanism shown in Listing 4 is all written as Smalltalk code. To avoid an infinite loop in this lookup logic, we apply a static type analysis to bind all messages during compilation. The result of the compilation is stored as an object known by the JIT compiler, which it will use when compiling send bytecodes, since they use this lookup code.

■ **Listing 4** The implementation of message dispatch in Bee/LMR

```
1  SendSite >> _dispatchOn: anObject
2      | cm nativeCode |
3      <specialABI: anObject -> regR>
4      <specialABI: self -> regA>
5      cm := anObject cachedLookup: selector.
6      cm == nil ifTrue: [^anObject doesNotUnderstandSelector: selector].
7      cm prepareForExecution.
8      nativeCode := cm nativeCode.
9      self when: anObject behavior use: nativeCode.
10     ^anObject _transferControlTo: nativeCode
```

## 5 Qualitative Evaluation of LMRs

This section first discusses LMRs and how they improve upon state-of-the-art VMs to solve the problems identified in Section 3.2. Then it details how the case studies can be solved using LMRs, and finally it discusses the trade-offs LMRs introduce.

### 5.1 How LMRs Improve Upon the State-of-the-Art VMs

We argue for the removal of the strict two-layer architecture and the separation between VM and application. Relying merely on object-oriented techniques for structuring programs allows us to resolve the problems identified in Section 3.2 as follows.

#### 5.1.1 Limited VM Observability (PG1)

As previously discussed, the strict two-layer architecture purposefully restricts what can be observed in the VM. Though, for our case study, it would be beneficial to more freely monitor not just the application but also the VM's behavior. With LMRs being *part of the application*, there is a more direct connection between application and VM code, which allows us to observe it in the same way as the application.

Programming systems implemented using LMRs also provide a more direct causal connections between application and VM code than traditional VMs, which can also be harnessed for instance to adapt more easily to unanticipated scenarios and helps us to improve applications in novel ways.

**Unified Programming Language**  One implicit benefit is that application developers do not necessarily have to learn and work with a different language. At least in the simplest realization of LMRs, the language used for the VM is the same as the one used for the application, with the small differences discussed in Section 4.2, which are more a set of conventions that can be learned on the fly.

**Unified Programming Tools**  Another benefit is that application programmers do not need to learn a new sets of tools, because their regular tools for code browsing, profiling, inspecting values, and debugging, are the same they normally use. However, changing a running system comes of course with the added risk of making changes that result in crashes. In our experience, this will result in a more careful experimentation, but does not discourage it.

#### 5.1.2 Separate VM Development Mode (PG2)

Since LMRs are essentially part of the application, they benefit from the same programming model as the application code, and support for Live Programming Environments. This also means there is no separate toolchain or build environment needed. Furthermore, all VM code is readily available to the developer. Benefiting from the same tooling also means that code is automatically compiled as the user accepts changes to VM code in the same way as it is done for application code. Thus, the application development mode is also the VM development mode.

**Incrementally Learnability**   Application developers using LMRs can browse the code of the different parts of the VM and inspect the relevant pieces, as they do with the code of the large application codebase. Similarly, instead of merely imagining how code behaves at run time, they can debug and observe it. For example, instead of imagining how or when the compiler triggers recompilation, it is possible to place a breakpoint in the compiler itself and debug the mechanism as it executes. Thus, it is possible to run the VM code step by step to understand how the pieces fit together, when and where it is needed.

#### 5.1.3 Long Edit-Compile-Run Feedback Loops for VM Components (PG3)

From the previous outlined benefits also follows that we have the same immediate feedback for VM code that we have for application code. This instant feedback reduces the time from ideas to experiments. In LMRs, compiling the VM does not require any different actions from the ones used for the application. The VM code is available in the development environment in exactly the same way as application code.

**Unified VM/Application Interface**   Another benefit of removing the architectural distinction between VM and application is that there is no need to devise a lower-level interface for passing objects to the VM from the application, nor an interface for using higher-level objects within the VM, since they use one uniform representation.

### 5.2 LMR-based Solution Approaches for the Case Studies

Based on the case studies described in Section 3.1, we now show how Bee/LMR enables us to solve the problems and overcomes the limitations identified in Section 3.2.

#### 5.2.1 Garbage-Collection Tuning (GCT)

Our first case study identified performance issues related to garbage collection. Specifically, opening the application from scratch can take over a minute and a significant amount of time is spent on allocating objects and garbage collection.

The first step is to understand where exactly time is spent. Since the GC is implemented as a library, it can be identified with the standard profiling tools together with the application code in context, which is often not possible in the same way on other VMs, because they try to make GC transparent and unobservable and tools present it separately. Because the GC is profiled the same as application code, the developer can see from a profile how much time was spent on allocating objects, how many garbage collections were triggered, and which parts of the GC, if any, are an area of concern.

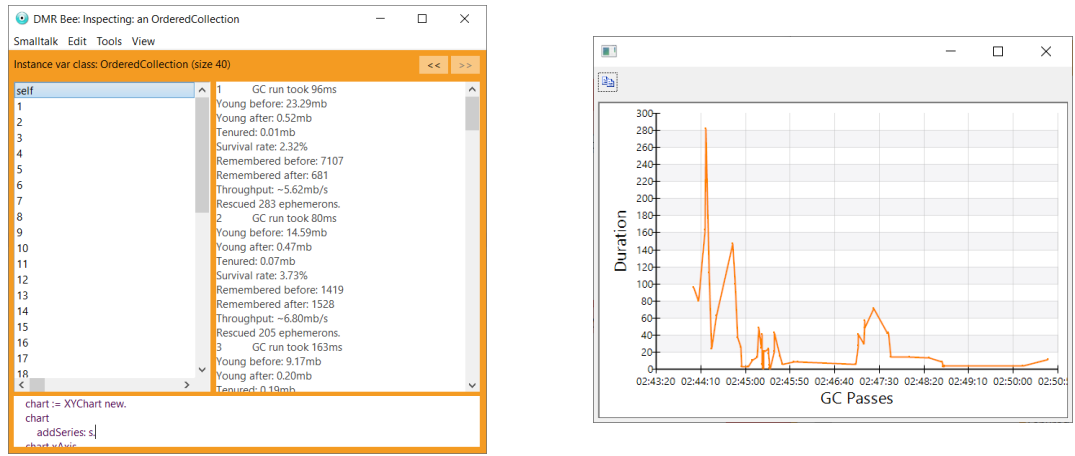

▪ **Figure 3** Debugging a garbage collector with standard application development tools. On the left, an inspector displays the garbage collection statistics of the live system, while on the right, a plot illustrates the time taken by the garbage collection process based on those statistics.

Once an area is identified as a performance bottleneck, we would want to collect more specific statistics from the GC to better understand why it spends its time there. Figure 3 shows the tools used during the case study, which are already known to the application programmer. Since the GC objects are directly available to the programmer, it becomes possible to directly modify them, while running, to collect statistics such as the heap sizes over time, object survival and tenuring rates, as well as the remembered set sizes. For instance, we can add an instance variables to `GenGCPass` to store the start time of the GC pass, and then modifying the code that triggers GC to store that value. In our case study, the developer used the obtained data to experiment with different Generational GC heap sizes and triggering heuristics while the system was running to observe and measure the changes in real time, having immediate feedback and data to improve the applications startup time. With this tuning, the application startup time was reduced by about 25 %.

#### 5.2.2 Recurring Recompilations by the JIT Compiler (JITC)

Similar to our GC tuning case study, profiling our application on Bee/LMR also helps identifying performance issues with the JIT compiler. Since the JIT compiler is merely another library, our application developer can use the profiler to identify why it takes too much time. In this case, the developer notices that the VM is repeatedly invalidating and recompiling code, because for the same class, it sees two different methods: the original one and the instance method added to an individual object. To provide a

**Removing the Barriers between Applications and Virtual Machines**

■ **Listing 5** The method NativizationEnvironment » nativeCodeFor: aBytecodeMethod obtains the native code needed to execute the method passed as argument. Adding the cache, adds support for instance-specific methods in the JIT compiler, preventing repeated recompilations.

```
1  NativizationEnvironment >> nativeCodeFor: aBytecodeMethod
2    cache at: aBytecodeMethod ifPresent: [:cached | ^cached].
3    nativeCode := methodNativizer nativeCodeFor: aBytecodeMethod.
4    (self shouldCache: aBytecodeMethod)
5       ifTrue: [cache at: aBytecodeMethod put: nativeCode].
6    ^result.
```

quick solution for the customer that found the problem in production, they try to modify the VM to enable the JIT compiler to cache the compiled code of both methods. Instead of triggering a recompilation when seeing an instance-specific method, they add a lookup in the compiled-code cache and compilation is only triggered if there is no matching cached entry.

The adapted code in Listing 5 implements this simple trick in the JIT compiler. While this method belongs to "VM code", it is a simple Smalltalk method and can be changed as any other method in the live programming environment. Furthermore, the developer does not need any specific knowledge of how the compilation works. They only needed to add the check of a dictionary and the caching of the native code in it. The tools, the language, and the concepts used are well known to the developer. Figure 4 shows a debugger halted at the point of assembling a bytecode. The liveness of the system allows developers to see the results instantly, without restarting the system, even coding in the debugger. Thus, this fix could be delivered to clients without requiring them to restart their application.

To summarize, when developers use an LMR to profile their applications, they have additional details on the operation of the JIT compiler. If the JIT compiler causes performance issues, it appears as any other part of the application, making it possible to identify opportunities for improvement. Thus, LMRs remove accidental complexity barriers and invite the developers to solve the problems more directly.

### 5.2.3 SIMD Optimizations (CompO)

The third case study aimed to add support for vectorized operations to speed up the floating-point arithmetics that are used in the simulation code of the application. Bee/LMR enables us to adapt the compiler to optimize the relevant code for us, which is not normally an option with state-of-the-art VMs. Since the compiler optimization can be application-specific, it is possible to support the bare minimum for the specific application instead of building a generic system that is suitable for a wide range of code structures. Not only does this give immediate benefits, but it also simplifies the problem significantly and allows a developer to incrementally support code patterns that are judged important in the specific application.

Listing 6 shows two implementations of element-wise addition of two float arrays. In this implementation, elements are added one by one. A SIMD-optimized version is shown in Listing 7. The programmer has to implement a compiler node that gets assembled to the desired `addps` instruction. The assembly can be debugged with a conventional Smalltalk debugger that includes native code information. As in the previous case studies, there is no need to perform tasks unrelated with the problem being solved, such as recompilation. The implementation of the assembly code is done while the system is running, without restarting, and can be changed and debugged as much as needed.

## 5.3 Discussion

By removing the strict separation between applications and VMs, we can benefit from more insight into the execution when utilizing our standard tools and changing the VM and runtime libraries becomes as direct and immediate as changing any application code. In the following, we will briefly discuss other benefits and drawbacks we found with this approach.

### 5.3.1 Additional Benefits

The flexibility LMRs bring provided us with a wider range of options when tackling problem scenarios and unanticipated change requests. Since LMRs reduce the length of the feedback cycle for VM changes by orders of magnitudes, experimentation and exploration takes much fewer time and effort. As noticed in the Section 5.2.2 when caching results of the JIT compiler, such kind of experimentation, and numerous changes can be done without particularly deep knowledge of the garbage collector, JIT

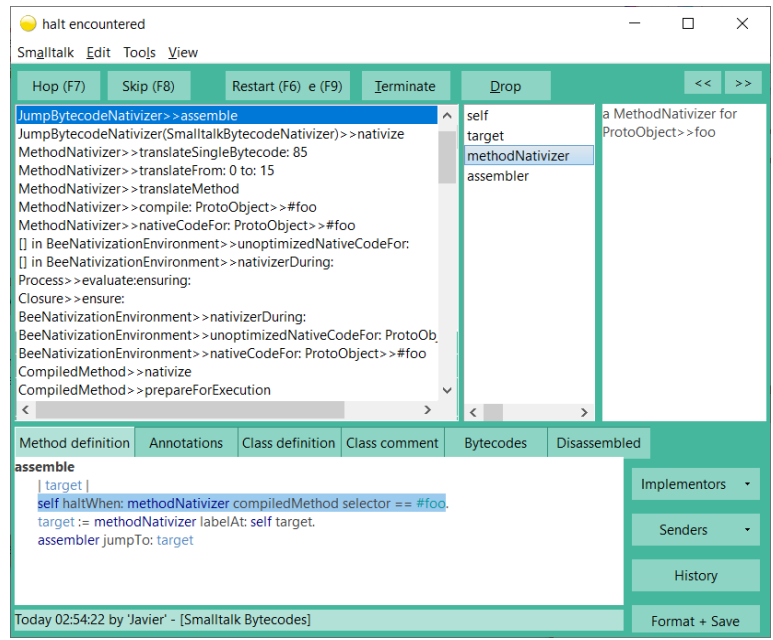

**Figure 4** A debugger displaying the template JIT-compiler assembling a jump

■ **Listing 6** The method FloatArray » += aFloatArray performs element-wise addition of the argument into the receiver. It invokes a naive implementation that performs one-by-one addition. _floatPlus:, _floatAt: and _floatAt:put: are metamessages that get evaluated at compile time and generate compiler nodes.

```
1  FloatArray >> += aFloatArray
2    (self checkAdditionArguments: aFloatArray) ifFalse:
3      [^self withIndexDo: [:f :idx | self atValid: idx put: f + aFloatArray]].
4    self basicPlus: aFloatArray
```

```
1  FloatArray >> basicPlus: aFloatArray
2    1 to: self size do: [:i | | a b |
3      a := self _floatAt: i.
4      b := aFloatArray _floatAt: i.
5      self at: i put: (a _floatPlus: b)]
```

```
1  Node >> _floatAt: indexNode
2    ^LoadNode base: self index: indexNode type: #Float32
```

```
1  Node >> _floatAt: indexNode put: valueNode
2    ^StoreNode base: self index: anONode value: valueNode type: #Float32
```

```
1  Node >> _floatPlus: rightNode
2    ^FloatPlusNode left: self right: rightNode.
```

```
1  X64CodeEmitter >> assembleFloatPlus: aFloatPlusNode
2    left := allocation at: aFloatPlusNode left.
3    right := allocation at: aFloatPlusNode right.
4    self assemble: 'addss' with: left with: right
```

■ **Listing 7** The SIMD version is quite similar, but it applies the operation to multiple data on each cycle. The number of iterations is divided by the amount of parallel additions, the type of stores and loads is adjusted to make compiler use appropriate SIMD registers and operand sizes. The assembly for the addition is changed to `addps`, a packed single-precision float addition.

```
1  FloatArray >> basicSimdPlus: aFloatArray
2    1 to: self simdSize do: [:i | | a b |
3      a := self _simdFloatAt: i.
4      b := aFloatArray _simdFloatAt: i.
5      self _simdFloatAt: i put: (a _simdFloatPlus: b)]
```

```
1  FloatArray >> simdSize
2    ^self size // self floatsPerSimdRegister
```

```
1  Node >> _simdFloatAt: indexNode
2    ^LoadNode base: self index: indexNode type: #SIMDFloat32
```

```
1  Node >> _simdFloatAt: indexNode put: valueNode
2    ^StoreNode base: self index: anONode value: valueNode type: #SIMDFloat32
```

```
1  Node >> _simdFloatPlus: rightNode
2    ^SIMDFloatPlusNode left: self right: rightNode
```

```
1  X64CodeEmitter >> assembleSIMDFloatPlus: aSIMDFloatPlusNode
2    left := allocation at: aSIMDFloatPlusNode left.
3    right := allocation at: aSIMDFloatPlusNode right.
4    self assemble: 'addps' with: left with: right
```

compiler, native code, or similar concepts that might be new to application developers. Often it is as simple as caching some result or collecting some data for getting a better understanding of what the application is doing.

For example, by looking at high-level native-code characteristics, one can learn about how the JIT compiler works and how optimization decisions of the JIT compiler affect the result. This makes it also possible to for instance detect methods being inlined or not being inlined. Depending on the context too few or too much inlining can result in worse performance. Since it is a live system, one can then tune heuristics and see the results without having to restart the system.

### 5.3.2 Importance of Liveness in LMRs

From the architectural perspective, it is possible to build runtime system that remove the separation between VM and application, without supporting liveness. Those runtime systems will likely also reduce the knowledge gap, improve causal connection between runtime and code, and enable application developers to better understand the runtime system. However, the instant feedback and observability of LMRs also turns the mere potential of these benefits into serendipitous encounters that will happen, which is key for the understanding of large codebases. Just because one may have access to the code, does not mean that one will look at it. In an LMR however, the code will be right there in the debugger or profiler one is using to diagnose a problem, and one can instantly experiment with it without any extra action to be taken. We believe this to be a qualitatively different situation than in a classic environment.

### 5.3.3 Drawbacks and Concerns Associated with LMRs

Removing the distinction between VM and application comes naturally with concerns about abandoning the benefits of this architecture. We will briefly discuss these and concerns such as software safety.

**Maintainability and Portability**   As previously discussed, the two-layer design is chosen by many VMs on purpose to ensure a strong separation of VMs and application. With VMs such as the Java Virtual Machine, this gives applications the opportunity to change between JVMs being confident that the application will continue to work. With LMRs, this strict separation and guarantee is not given anymore. However, we would argue that in some situations this is a trade-off worth making.

Considering that many large applications would use the technique of *vendoring* frameworks they rely on into their own codebase, an LMR is essentially the practice of vendoring the runtime library into the application. Vendoring is typically done to gain more stability and be able to fully control a framework or library. On the flip side, this comes at the cost of maintaining this fork and upstreaming changes into the original. Arguably, the same is true for LMRs and as with vendored frameworks, it gives all the flexibility and cost of doing so. However, one may always decide not to change anything, which will typically make it relatively simple to stay current with any changes made upstream.

**Safety** When creating a virtual machine (VM) for a programming language, software safety is a crucial concern. We argue that moving the code between architectural layers does change the situation only superficially and any concrete threats are specific to the programming language of the LMR. As such, there is no general answer and whether an LMR introduces new attack surfaces depends on the language in question.

For Bee/LMR, the situation is not ideal from the start. As a Smalltalk, there are many opportunities to attack the system by evaluating arbitrary code at run time or by using the #become: operation to arbitrarily swap objects with each other. Languages such as Java and JavaScript typically have mechanisms to restrict what can be changed. For instance, Java's module system is also able to prevent reflective accesses. JavaScript has the freeze() method that allows it to make objects immutable. Such mechanisms could in theory be used to provide some form of protection. However, having a JIT compiler, its assembler accessible, and the ability to access arbitrary memory in theory allows the execution of arbitrary code and as such would require careful design to prevent an attacker to gain access to this capability. Type and capability systems, for instance using mirrors [7], could facilitate a suitable system design.

**Stability** From a practical standpoint, developers have flexibility in choosing their preferred workflow. Changing the system while it is running is risky, as even a minor mistake can cause the system to crash. In Bee/LMR, there are no special crash avoidance mechanisms than the ones given by the language itself. That includes safe indexing of arrays and mechinsms such as `doesNotUnderstand`, which let programmers catch and fix errors as they show up. It is possible and common practice to save the image right before applying such changes. Additionally, if a crash does occur, a remote debugger pops up, allowing to inspect the frozen image before terminating the process, to aid understanding the cause of the crash. Alternatively, since Bee/LMR can be edited as code offline, one can also modify it and then bootstrap it, which allows applying more complex changes that might not be safe in a running system. In practice, developers often start with the live workflow and switch to the latter once they found the boundaries of what is possible in a live system. The same holds for designing the overall system. Often we find ourselves to chose more conservative designs, which facilitates live updates and reduces the risk of stability issues.

To make this more concrete, let us consider the garbage collector a bit closer. Since the GC is a crucial component that needs to be work at all time, we need to ensure *always working objects*, and prevent *lost writes* and *lost objects*. This restricts design choices for GC algorithms but allows us to build a more reliable system.

*Always working objects.* As the GC code depends on the interaction of common objects, during GC it is not possible to apply GC algorithms that temporarily overwrite object data, such as those that install forwarding pointers into object headers or that thread reference pointers. With such algorithms, if the GC tried to send a message to a temporarily overwritten object it could crash because it would not find the expected meta-information of the object such as its class or its size.

*Lost Writes.* The garbage collector has to avoid moving objects that might change during garbage collection. During GC, objects are typically moved (copied) to another area when being reached for the first time when tracing the object graph. As the GC continues tracing, it can find more references to the moved objects and update them with the new addresses. If the GC modifies a moving object, it might cause inconsistencies because it may be impractical to determine whether an object is the new copy or the old one, and during GC some objects might still see the original version instead of the modified one because they have not been traced yet.

*Lost Objects.* The garbage collector switches the allocation area when it starts, so that objects needed for GC do not get mixed with the rest and get discarded as soon as GC finishes. This means that no new objects created during GC survive after GC ends. In particular, it is not allowed to JIT-compile methods during GC for that reason. The system must assure that any code required for executing GC has already been compiled before the GC starts.

### 5.3.4 Metamodel Dynamicity and System Scalability

Our LMR implementation does not expand on the metaobject protocol to allow changes such as modifying object layout format on-the-fly. While in principle, one can treat all objects, or rather the programs memory as bits, and run a script over them to do such updates, Bee does not provide any special support or API to make such changes convenient. Small changes like modifying the meaning of a free bit in object headers may be explored while running the program. Bigger changes, on the other hand, usually get done by modifying source files and re-bootstrapping the system.

When considering the scalability of an LMR, one may be concerned about the support of large object heaps as well as the support for large codebases. Since we use the garbage-first GC design [12], we have not noticed any scalability issues. Garbage-first is designed for large heaps and to give soft real-time guarantees on garbage collection pauses.

When it comes to large codebases, the Smalltalk approach of an image-based system, that contains the code, scales quite naturally. Since code is compiled at run time and development time one method at a time and on-demand, the growing codebase has not been a notable concern, even with our 1.1 million lines of code application.

## 6 Quantitative Evaluation

To assure the viability of our approach, we run a series of performance tests that provide a lower bound to the performance of a LMR-based systems. Practical performance is not a problem for LMRs. Bee/LMR is used daily by a team of four developers and is deployed to run our product in production. It receives two major releases per year, with monthly minor releases for specific customers.

### 6.1 Performance Evaluation

The goal of this evaluation is to show that LMRs can reach the performance levels of traditional VM-based systems. We compare Bee/LMR against the following systems:

**Pharo/Cog** Pharo is a Smalltalk dialect which runs on top of OpenSmalltalk VM, the most widely used Smalltalk VM. We run OpenSmalltalk run in dual JIT/interpreter mode (Cog). OpenSmalltalk is written in Slang, transpiled to C and compiled with a standard C compiler.

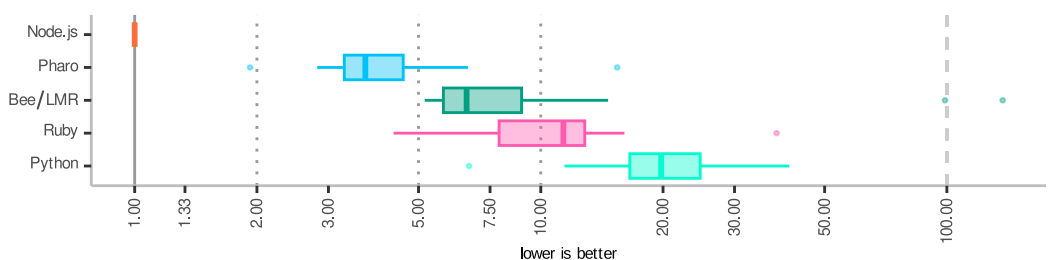

**Figure 5** Bee/LMR benchmark times compared to other dynamic object-oriented systems.

**Python/CPython** The standard Python VM (v3.10.7) which is written in C and uses an interpreter.

**Ruby/MRI** The standard Ruby VM (v3.0.4p208), also written in C.

**JavaScript/V8** The JavaScript engine behind Node.js (v18.16.0), written in C++, that includes an interpreter and several optimizing JIT-compiled stages.

We use the *Are We Fast Yet* benchmarks which include 9 micro and 5 macro benchmarks [20]. They are designed to compare performance across different language implementations and thus were easy to adapt to Bee.

The benchmarks were run on a machine with a 2.8Ghz 4-core Core i7 7700HQ with hyper-threading and 16 GB of memory. The operating system is a 64-bit Ubuntu 22.10. Bee is run through Wine compatibility layer, as it only supports Windows. All the implementations are 64-bits.

We measure 100 iterations of each benchmark, and in the case of Node.js 3000 to minimize noise from late JIT compilation. For each benchmark, we take the median of the measurements and report the summary as a standard boxplot in Figure 5. The results are normalized to Node.js, which is the fastest system.

Overall Bee/LMR is slightly slower than Pharo/Cog. This is expected, since the Cog VM has been optimized for a much longer period than Bee/LMR. For example, string copying in Bee is done byte-by-byte, checking bounds at each character. Bee is around one order of magnitude faster than Ruby/MRI and Python/CPython interpreters. SinceBee/LMR uses a template-based JIT compiler [3], there is a lot of room for improving its performance with classic compiler optimizations, when comparing with Node.js and Pharo. However, since CPython and MRI are classic bytecode interpreters without JIT compilation, Bee/LMR outperforms them roughly at the level one would expect from a template-based compiler. Python is known to be one of the slower interpreters, which newer versions starting with Python 3.11 aim to fix.[4]

---

[4] https://docs.python.org/3.11/whatsnew/3.11.html#faster-cpython Last access: 2023-10-27

**Removing the Barriers between Applications and Virtual Machines**## 6.2 Runtime Implementation Size

We evaluate the size of Bee/LMR by counting the lines of code of different parts of the system associated with VM components: JIT compiler, GC and built-in functions. We compare that metric against the other systems to give an idea of the difference. Bee/LMR is significantly smaller, because it is much more specialized and does not support the same variety of operating systems, processor architectures, and usage scenarios.

| Subsystem | LoC |
|---|---|
| GC | 1,689 |
| Baseline JIT | 3,329 |
| Optimizing Compiler | 5,567 |
| Intel Assembler | 8,868 |
| Built-ins | 2,604 |
| **Total** | **22,057** |

**(a)** Lines of codes of Bee/LMR subsystems

| VM | LoC | |
|---|---|---|
| V8 (JavaScript) | 1,111,919 | C++ |
| CPython | 245,538 | C |
| MRI (Ruby) | 210,120 | C |
| OpenSmalltalk | 124,741 | Smalltalk |
| **Bee/LMR** | **22,057** | Smalltalk |

**(b)** Lines of codes of different dynamic, object-oriented virtual machines, compared to Bee/LMR

■ **Figure 6** Size of the LMR module in lines of code

## 7 Conclusions

We propose Live Metacircular Runtimes (LMRs), a runtime design that combines VM implementation and application, replacing the traditional architecture that separates them into layers. Our approach uses basic object-oriented techniques instead to ensure for instance encapsulation and to enable changes to the runtime systems at run time, the same as normal application code.

In this paper, we use case studies on tuning the garbage collector, changing the just-in-time compiler to avoid unnecessary recompilations, and adding support for vector instructions to the compiler to argue for the benefits of removing the architectural separation to enable shorted feedback cycles and better understanding of the runtime system by application developers.

Bee/LMR is our implementation of Bee Smalltalk that runs on top of an LMR, instead of a traditional VM. It is used in production to run a 1.1 million lines of code application. This LMR-based system has allowed application developers to incrementally understand VM components as needed, and even to modify them when required as argued with our case studies. Specifically, we show that this approach avoids the limited VM observability, the separate VM development mode, and the long edit-compile-run cycles for the traditional two-layer architecture.

In future work, we want to explore how to improve performance without sacrificing the ability to change the runtime at run time. While compilers such as Graal demonstrate that compilers in high-level languages can produce state-of-the-art performance, our design of the object layout and garbage collector made careful decisions to preserve a working system at all times. There are similar circular dependencies in the just-in-time compiler, which means that a change at run time could break it. This could be mitigated by supporting multiple versions of such subsystems, where the old working version remains available as a fallback.

**Acknowledgements** This work was supported by European H2020-MSCA-RISE-2017 project "Behavioural Application Program Interfaces (BEHAPI)" (ID 778233), by a grant from the Engineering and Physical Sciences Research Council (EP/V007165/1) and a Royal Society Industry Fellowship (INF\R1\211001).

## A  Other Uses

While not part of the in-depth case studies for the lack of space, this appendix provides a description of other use cases that were discovered during real-life usage of our LMR-based system.

### A.1  Memory Leaks Detection

The presence of a live system allowed for novel uses of the memory management runtime that are typically not practical with static VMs. The GC can be modified lively and harnessed in unanticipated ways.

In Bee/LMR, we harnessed the GC to find out memory leaks in an HTTP request server. Bee includes a generational GC made of an `eden` space and two flipping `from` and `to` areas where objects go before moving to old area. Because the GC is integrated as a library, it is possible to track live objects in particular spaces, so finding leaked objects can be done by the following actions:

1. Before handling a request, trigger a generational GC to clean `eden` space.
2. Disable GC during the handling of the request, so that all objects get allocated in the `eden`, making `eden` grow if necessary during the request.
3. After the request gets handled, trigger a generational GC so that live `eden` objects get moved to the `from` space.
4. Using the GC API, iterate through the objects in the `from` space adding them to a weak collection. The contents of this collection is a superset of the leaked objects. It may still retain objects that were pointed from remembered set but not really alive.
5. Run a full GC to nil out the entries of the collection that are unreachable when tracing the full object graph. The contents of this set is now the set of leaked objects.

The code for performing these actions fits in one method, and was added to the memory manager and tested with immediate feedback. The API receives a block closure and returns the objects that survived the evaluation of that block. The code is shown in Listing 8.

As in other situations, this API is mostly useful at development time rather than at deployment time. In the traditional VM scenario, implementing this would have

■ **Listing 8** Memory»objectsSurviving: aClosure method evaluates the closure passed as a parameter and returns the objects that were leaked. `HttpWorker»processRequest: anHttpRequest` uses that API to graphically show the result.

```
1  Memory >> objectsSurviving: aClosure
2      | set finalizable |
3      set := WeakIdentitySet new.
4      self collectYoung; disableGC.
5      aClosure value.
6      self enableGC; collectYoung.
7      fromSpace objectsDo: [:o | set add: o].
8      self collect; collect.
9      ^set
```

```
1  HttpWorker >> processRequest: anHttpRequest
2      leaked := Smalltalk memory
3          objectsSurviving: [ self doProcessRequest: anHttpRequest].
4      leaked inspect.
```

required passing through all the barriers detailed in Section 3.2, implementing the change in the VM, recompiling it restarting the application with the patched VM and then trying the experiment. In the LMR, the experiment was done without delays. Actually, the first try saved the results into a standard not-weak collection, and let us discover that we needed an extra filtering of objects through full GC. In the traditional VM case, that would have required another recompilation and restart step that was not needed in the LMR.

### A.2 Implementation of Code Coverage of Tests

Bee contains tools for analyzing code coverage of tests. These tools were based on instrumentation of compiled methods. While useful for most of the typical scenarios, the instrumentation approach was too slow to be executed on all the 15000 tests of the system that stress the 1.1 million lines of code of the application.

This problem was solved by accessing JIT-compiler information from the application, instead of using instrumentation: As Bee/LMR only executes code by JIT-compiling methods, if the JIT code cache is cleared before running tests, and the size of that cache is big enough, then it is possible to determine which methods have been executed by just checking whether they got added to the code cache. This information is readily available in the system, and can be collected with a script like shown in Listing 9.

The required change was implemented by application developers in the unit test library, and allowed them to obtain coverage statistics without paying performance penalties.

■ **Listing 9** TestSuite » coverage method clears the code cache, runs a test suite and finally counts how many of the methods in the system have been executed by checking how many have been assigned native code.

```
1  TestSuite >> coverage
2      | methods executed |
3      Smalltalk clearCodeCache.
4      self run.
5      methods := CompiledMethod allInstances.
6      executed := methods count: #hasNativeCode.
7      ^executed / methods size
```

■ **Listing 10** a BitsAt node contains a pair of left and right sub-nodes. The left one can be anything, the right one is a constant pointing to a BitField object. The optimized method accesses the constant and generates optimized code (a bit and followed by a bit shift).

```
1  BitsAt >> optimized
2      | bitfield and shift |
3      bitfield := right value.
4      self assert: bitfield class == BitField.
5      and := BitAnd left: left right: bitfield mask.
6      shift := BitShift left: and right: bitfield shift.
7      ^shift
```

### A.3  Other Optimizations

At some point our simulation application incorporated a BitField type that simplified working with bit fields. The type allowed extracting bits of an integer as if it were a C bit field, and also writing bits back. The usage is simple, the client creates a bit field with something like flags := BitField bits: 4 to: 6 and then they can extract and write the bits back with for example flags bitsAt: field and flags bitsAt: field put: aValue.

After some time, it was discovered that this simple abstraction was causing a small performance penalty. As the compiler could not prove that bit fields were immutable, the operations for *shifts* and *ands* were using generic code with fallback cases for non-integer types. Adding general immutability support to the compiler was out of the scope, but instead of removing the abstraction it was possible to hand tune Bee compiler to incorporate an ad-hoc optimization. With this optimization, when the compiler sees a bitsAt: message sent to a bit-field object, it assumes it is immutable and generates optimized code for the specific size of the bit field, removing all performance penalties.

The implementation of this optimization involved two steps: the first one was adding bitsAt: and bitsAt:put: to a list of special messages in the compiler. When seeing such a message, the compiler converts the MessageSend node to a specialized BitsAt node, a subclass of BinaryMath nodes. The second step was the optimization itself. For that, the `BitsAt` node implements the optimized method, that in turn converts itself into a series of *shift* and *and* nodes, as shown in Listing 10.

As the optimization allowed zero-cost bit fields, this abstraction was then incorporated back into the VM code, simplifying the code that accesses bit fields stored in compiled methods and classes.

**Removing the Barriers between Applications and Virtual Machines**

## About the authors

**Javier E. Pimás** is the author of Bee/LMR. He works at Quorum Software while also pursuing a PhD in Computer Science at Buenos Aires University, where he occasionally does teaching. He is a fan of high-level low-level programming, and has been successfully mixing Smalltalk and assembly code for more than a decade. Within Bee Smalltalk, SqueakNOS and Powerlang projects he has been trying to make live-programming of Virtual Machines more practical to application programmers. Contact Javier at jpimas@dc.uba.ar.  0009-0008-4668-2306

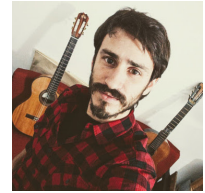

**Stefan Marr** is a Senior Lecturer and Royal Society Industry Fellow at the University of Kent. He is interested in a wide range of language implementation techniques to improve the performance of modern languages and to enable better tools that allow developers to better understand and maintain their applications' code. Contact Stefan at s.marr@kent.ac.uk.  0000-0001-9059-5180

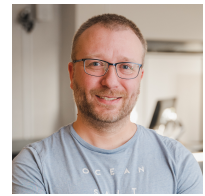

**Diego Garbervetsky** is an Associate Professor at the Computer Science Department, School of Sciences, University of Buenos Aires. He is also a Researcher at the ICC/CONICET, and also Director of the Institute of Research in Computer Sciences(ICC).

He works on static analysis techniques aimed at Java-like programs and Smart Contracts, automated program verification, program understanding and validation. Contact Diego at diegog@dc.uba.ar.  0000-0003-4180-7196

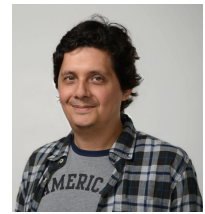